\def\bey{\begin{eqnarray}}
\def\eey{\end{eqnarray}}
\def\be{\begin{equation}}
\def\ee{\end{equation}}
\def\gm{\gamma}

\def\Ld{\Lambda}

\def\sg{\sigma}
\def\Sg{\Sigma}

\def\om{\omega}
\def\r{\rho}

\def\bt{\beta}

\def\pp{\partial}
\def\pp{\partial}

\def\nnb{\nonumber}
\documentstyle[11pt,epsfig]{article}
\title{Roles of isoscalar hyperons in probing the density dependence of the nuclear symmetry energy}
\author{ W. Z. Jiang\footnote{Email: jiangwz02@hotmail.com, Also at  Center of Theoretical Nuclear Physics,
 National  Laboratory of Heavy Ion Accelerator, Lanzhou 730000,China}\\
   Shanghai Institute of Applied Physics,\\
  Chinese Academy of Sciences,Shanghai 201800,China
 }
\date{}
\bigskip
\begin{document}
\maketitle
\baselineskip 20.6pt
\begin{abstract}
\baselineskip 18.0pt The role of the isoscalar hyperon $\Ld$ in
probing the density dependence of the nuclear symmetry energy is
studied in  multi-$\Ld$ hypernuclei, hyperon-rich matter and
neutron stars in relativistic  models. Relationships between the
properties of three types of objects and the neutron thickness in
$^{208}$Pb are established  with respect to the
isoscalar-isovector coupling that modifies the density dependence
of the symmetry energy. The exotic isotopes far from the neutron
drip line can be stabilized by filling in  considerable $\Ld$
hyperons.  The difference of the binding energy of multi-$\Ld$
hypernuclei from different models is attributed to different
symmetry energies.  The isovector potential together with the
neutron thickness in  multi-$\Ld$ hypernuclei investigated is very
sensitive to the isoscalar-isovector coupling.  The large
sensitivity of the $\Ld$ hyperon fraction to the
isoscalar-isovector coupling occurs at about  $2\sim3\r_0$ in beta
equilibrated hyperon-rich matter. In neutron stars with
hyperonization, an on-off effect with respect to the
isoscalar-isovector coupling exists for the neutron star radius.

\end{abstract}



\newpage

The symmetry energy is pervasively related to the descriptions  of
nuclear structures and reactions in laboratories and
astrophysics\cite{bethe,bomb,stein3}. However, the density
dependence of the symmetric energy is still poorly
known\cite{brown,li} to date. Therefore,  theoretical
uncertainties  exist in the understanding of a large number of
properties of neutron stars and  heavy nuclei\cite{horo1,horo2},
nuclei far from the $\bt$ stability\cite{todd,weiyb,jiang1}, heavy
ion collisions\cite{li,chenlw} and consequently probable
hadron-quark phase transition\cite{ditoro}, etc. Many factors that
influence the isovector potential directly or indirectly can
modify the behavior of the density dependence of the symmetry
energy. Recently the density dependence of the symmetry energy has
been extensively explored through the inclusion of the
isoscalar-isovector coupling terms in relativistic mean field
(RMF) models\cite{horo1,horo2,todd,horo4}. These new terms enable
one modify the neutron skin of heavy nuclei without changing the
charge radius and a variety of ground-state properties that are
well constrained experimentally. On the other hand,  the charge
density distributions for some light exotic nuclei near drip lines
that are not determined yet are sensitive to the
isoscalar-isovector couplings\cite{jiang1,jiang2}.

Hypernuclei that are produced through  strangeness transfer
reactions mostly contain only one hyperon, and  just a few double
$\Ld$ hypernuclear events were observed\cite{aoki,ahn,taka}.
Through heavy-ion collisions it may possibly create an environment
which is favorable to the formation of metastable exotic
multihypernuclear objects (MEMOs) such as multi-$\Ld$ hypernuclei.
Schaffner et.al. showed that multi-$\Ld$ hypernuclei were more
strongly bound than normal nuclei using the RMF
model\cite{schf1,schf2}. Recently, Jevorsek et.al. set limits on
the existence and properties of exotic objects such as strangelets
and MEMOs, and provided limits as low as $10^{-7}$ for M/Z up to
120 through a reanalysis of related data\cite{jev}. In bulk
matter, strangeness can be formed by virtue of the strong
interaction, and hyperons may be important constituents of neutron
stars. Many important effects of hyperonization in neutron stars
have been found in the past (for a review,see\cite{glend}). For
instance, hyperonization  reduces the maximum mass of neutron
stars as much as 3/4$M_\odot$\cite{glend85,glend91}. Schaffner
et.al. found that a phase transition to hyperon matter may even
possibly occur in neutron stars\cite{schf3}.  In this work, it is
the aim  to constrain the density dependence of the symmetry
energy through the investigation of properties of multi-$\Ld$
hypernuclei, hyperon-rich matter and neutron stars with
hyperonization. We will perform the investigation in RMF models
considering the isoscalar-isovector coupling terms that soften the
symmetry energy at large densities.

The modification in MEMOs and hyperon-rich matter due to the
isoscalar-isovector couplings is usually considered to be small
due to the relatively weak meson-hyperon coupling. As pointed out
in Refs.\cite{schf1,schf2} that the MEMOs may reach a very high
baryon density ($2.5\sim3\r_0$) like strangelets. In hyperon-rich
matter, the appearance of $\Ld$ hyperons occurs at about twice
normal density. Since the higher density relates to the larger
isoscalar potentials (or fields), the isoscalar-isovector
couplings will induce large modifications.  In hyperon-rich
matter, we will focus on the relation between the $\Ld$ fraction
and the isoscalar-isovector coupling. As the determination of
properties of neutron stars with  hyperonization needs the
equation of state of hyperon-rich matter,  we may search for the
observable consequence relating to  the density dependence of the
symmetry energy. As examples of MEMOs, we will consider
multi-$\Ld$ hypernuclei based on the Ca isotope far off the normal
neutron drip line at the fixed isospin. In this way, we can
observe the extension of the drip line, and  explore  the density
dependence of the symmetry energy by shifting the isoscalar field.

In this study, additional strange mesons $\sg^*$ (i.e. $f_0$,
975MeV) and $\phi$ (1020MeV)  are also included in  RMF models to
describe the strong $\Ld\Ld$ attraction as in
Refs.\cite{schf2,schf3,schf4}.  The effective Lagrangian density
is given as follows
 \bey
 {\cal L}&=&
{\overline\psi}_B[i\gm_{\mu}\partial^{\mu}-M_B+g_{\sg B}\sg-g_{\om
B} \gm_{\mu}\om^{\mu}-g_{\r B}\gm_\mu \tau_3 b_0^\mu\nnb\\
 & &+\frac{f_{\om B}}{2M_N}\sg_{\mu\nu}\pp^\nu \om_0^\mu
  -e\frac{1}{2}(1+\tau_c)\gm_\mu A^\mu]\psi_B-U(\sg,\om^\mu, b_0^\mu)\nnb\\
&& +\frac{1}{2}
(\partial_{\mu}\sg\partial^{\mu}\sg-m_{\sg}^{2}\sg^{2})-
\frac{1}{4}F_{\mu\nu}F^{\mu\nu}+
      \frac{1}{2}m_{\om}^{2}\om_{\mu}\om^{\mu} \nnb\\
      &  &
    - \frac{1}{4}B_{\mu\nu} B^{\mu\nu}+
      \frac{1}{2}m_{\r}^{2} b_{0\mu} b_0^{\mu}-\frac{1}{4}A_{\mu\nu}
      A^{\mu\nu} +{\cal L}_Y
\label{eq:lag1}   \eey
 and ${\cal L}_Y$ is for the strange
meson-hyperon interactions and free fields of strange mesons
 \bey
 {\cal L}_Y&=&
{\overline\psi}_Y[g_{{\sg^*}Y}\sg^* - g_{\phi Y}
\gm_{\mu}\phi^{\mu}]\psi_Y+\frac{1}{2}
(\partial_{\mu}\sg^*\partial^{\mu}\sg^*-m_{\sg^*}^{2}{\sg^*}^{2})\nnb\\
&& - \frac{1}{4}(\pp^\mu \phi^\nu -\pp^\nu \phi^\mu )(\pp_\mu
\phi_\nu -\pp_\nu \phi_\mu )+
      \frac{1}{2}m_{\phi}^{2}\phi_{\mu}\phi^{\mu}
\label{eq:lag2}   \eey
 where $\psi_B,\sg,\om$, and  $b_0$  are the fields of
the baryon,  scalar, vector, and charge-neutral isovector-vector
mesons, with their masses $M_B, m_\sg,m_\om$, and $m_\r$,
respectively. $A_\mu$ is the field of the photon.
$g_{iB}(i=\sg,\om,\r)$ and $f_{\om B}$ are the corresponding
meson-baryon couplings. $\tau_3$ is the  third component of
isospin Pauli matrix for nucleons and $\tau_3=0$ for the $\Ld$
hyperon.  $\tau_c$ is a constant relating to the baryon charge.
$F_{\mu\nu}$, $ B_{\mu\nu}$, and $A_{\mu\nu}$ are the strength
tensors of the $\om$, $\r$ mesons and the photon, respectively \be
F_{\mu\nu}=\pp_\mu \om_\nu -\pp_\nu \om_\mu,\hbox{ }
B_{\mu\nu}=\pp_\mu b_{0\nu} -\pp_\nu b_{0\mu},\hbox{ }
A_{\mu\nu}=\pp_\mu A_{\nu} -\pp_\nu A_{\mu}\ee The
self-interacting terms of $\sigma,\om$-mesons and  the
isoscalar-isovector ones are in the following
 \bey
 U(\sg,\om^\mu, b_0^\mu)&=&\frac{1}{3}g_2\sg^3+\frac{1}{4}g_3\sg^4
 -\frac{1}{4}c_3(\om_\mu\om^\mu)^2\nnb\\
 &&-4g^2_{\r N} (\Ld_s g_{\sg N}^2\sg^2+\Ld_{\rm v} g_{\om N}^2
 \om_\mu\om^\mu)b_{0\mu}b_0^\mu \label{equ1}
 \eey
For the hyperon-rich matter, the chemical equilibration is
established on weak interactions of baryons and leptons as in
Ref.\cite{glend85,prak}. We study  chemically equilibrated and
charge neutral hyperon-rich matter including baryons $(N,\Ld,\Xi)$
and leptons $(e,\mu)$ without $\Sg$ hyperons, since $\Sg$ hyperons
do not appear\cite{schf3,schf4} according to the fact that the
isoscalar potential changes sign in the nuclear interior and
becomes repulsive based on recent $\Sg^-$ atomic data
\cite{mares}. For the finite system, we consider the spherical
case. The solution detail can be referred to
Refs.\cite{serot,ring}, and it is not repeated here.  The pairing
correlation of nucleons in nuclei is considered in the BCS
approximation as in Refs.\cite{jiang2}.

We perform calculations with the NL3 \cite{nl3} and
S271\cite{horo1} parameter sets. These two models share the same
binding energy per nucleon (16.24MeV) and incompressibility
(271MeV) at the same Fermi momentum $k_F=1.30$fm$^{-1}$
($\rho_0=0.148$fm$^{-3}$).  The symmetry energy is given
by\cite{horo2} \be a_4=\frac{k_F^2}{6E^*_F}+\frac{g_\r^2}{3\pi^2}
\frac{k_F^3}{{m_\r^*}^2}\label{sym}\ee with
$E_F^*=\sqrt{k_F^2+{M^*_N}^2}$ and ${m^*_\r}^2=m_\r^2+8g_{\r
N}^2(g_{\sg N}^2\Ld_{\rm s} \sg^2+g_{\om N}^2\Ld_{\rm v}
\om_0^2)$. The density dependence of the symmetry energy is
different between two models because of different $g_{\r N}$,
$m^*_\r$ and $M^*_N$ ($M^*_N=0.59M_N$ for the NL3 and
$M^*_N=0.7M_N$ for the S271).

The $\om$ and $\rho$ meson couplings to the hyperons are given by
the SU(3) relations:  $g_{\om\Ld}=2/3g_{\om N}=2g_{\om\Xi}$,
$g_{\r\Ld}=0$ and $g_{\r\Xi}=g_{\r N}$. The $\sg$ meson coupling
to the hyperons is determined by  reasonable hyperon potentials in
nuclear matter: $U^{(N)}_\Ld=-30$MeV and $U^{(N)}_\Xi=-20$MeV. The
coupling constants  $g_{\phi\Ld}$ and  $g_{\phi\Xi}$  are taken to
satisfy the SU(6) relation: $g_{\phi\Xi}/g_{\om
N}=2g_{\phi\Ld}/g_{\om N}=-2\sqrt{2}/3$, and the $g_{{\sg^*}\Ld}$
is fitted to improve the $\Ld\Ld$ interaction matrix element
$\Delta B_{\Ld\Ld}$\cite{schf2}. In Ref.\cite{schf3}, the
 $g_{{\sg^*}\Ld}$ is taken in a range close to the
$g_{\sg N}$. We take the ratio of the scalar coupling constant to
be $g_{{\sg^*}\Ld}/g_{\sg N}=0.76$ in NL3 with a potential for the
$\Ld$ hyperon in $\Xi$ matter $U^{(\Xi)}=-55.6 $MeV. In S271, the
same potential is taken to obtain the ratio $g_{{\sg^*}\Ld}/g_{\sg
N}=0.77$. In this way, the $\Delta B_{\Ld\Ld}$ of
$^{10}_{\Ld\Ld}$Be is improved from 0.38 MeV to 2.2 MeV in NL3,
and from 0.2 MeV to 3.9 MeV in S271, much close to the extracted
value $4\sim5$ MeV. The $g_{{\sg^*}\Xi}$ is determined by  the
$\Xi$ hyperon  potential in $\Xi$ matter $U^{(\Xi)}_\Xi=-40
$MeV\cite{schf2}. The $\om NN$ tensor coupling is vanishing, and
the $\om \Ld\Ld$ tensor coupling is small but adjusted to simulate
the vanishing spin-orbit splitting for the $\Ld$ hyperon observed
in $^{16}_\Ld$O.

For simplicity, we perform calculations for various
isoscalar-isovector coupling $\Ld_{\rm v}$'s, and  the $\Ld_s$ is
set as zero. For a given coupling $\Ld_{\rm v}$, we follow
Ref.\cite{horo1,todd} to readjust the $\r NN$ coupling constant
$g_{\r N}$ so as to keep an average symmetry energy fixed as 25.7
at $k_F=1.15$ fm$^{-1}$. In doing so, it was found in
Ref.\cite{horo1} that the binding energy of $^{208}$Pb is nearly
unchanged for various $\Ld_{\rm v}$'s.  Below, we will see how the
total binding energy of multi-$\Ld$ hypernuclei  is modified by
the  $\Ld_{\rm v}$.

\begin{figure}[thb]
\begin{center}
\vspace*{-25mm} \epsfig{file=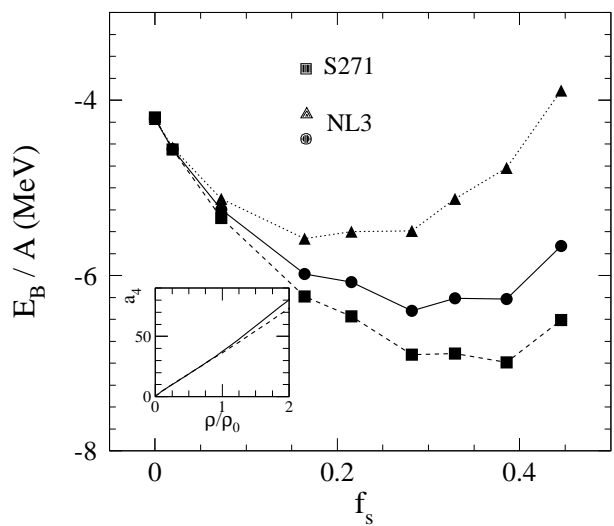,height=10.0cm,width=10.0cm}
 \end{center}
\caption{Binding energy per baryon for various $^{102}$Ca+$\Ld$'s
systems as a function of the strangeness fraction $f_s=n_\Ld/n_B$
at $\Ld_{\rm v}=0$. The symbols from left correspond to the
hyperon number 0, 2, 8, 20, 28, 40, 50, 64, and 82, respectively.
Except for results marked by the triangles, the strange mesons are
considered. The solid and dashed curves in the inset are for the
symmetry energy in NL3 and S271, respectively.
 \label{fig1}}
\end{figure}

The binding energy per baryon for various $^{102}$Ca+$\Ld$'s
systems is displayed in Fig.\ref{fig1}. Metastable minima appears
with the rise of  the $\Ld$ hyperon number $n_\Ld$.  The
additional $\Ld\Ld$ attraction, simulated by the strange mesons,
produces much larger binding energy at larger $f_s$ and shifts the
minima outwards.  The difference of  binding energies of
multi-$\Ld$ hypernuclei  in NL3 and in S271 enlarges explicitly
with the rise of $n_\Ld$.   Since both models share the same
saturation properties and take the same potential for the $\Ld$
hyperon,  this enlargement is attributed to the different density
dependence of the symmetry energy. With the rise of $n_\Ld$, the
central density of multi-$\Ld$ hypernuclei  increases. For
instance, the central density in $^{102}$Ca$+50\Ld$  is about
1.43$\rho_0$ in NL3 which is much larger than  1.06$\rho_0$ in
$^{102}$Ca.  The more different symmetry energy at the larger
density domain, shown in the inset in Fig.\ref{fig1}, thus
accounts for the large difference of the binding energy at larger
$f_s$.

\begin{figure}[thb]
\begin{center}
\vspace*{-25mm} \epsfig{file=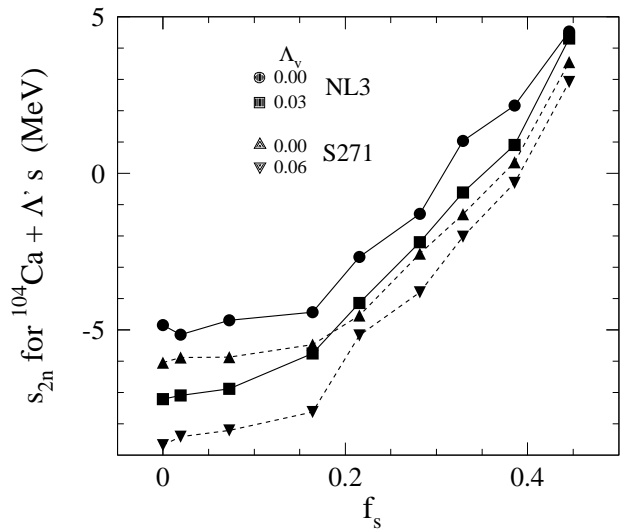,height=10.0cm,width=10.0cm}
 \end{center}
\caption{ Two-neutron separation energy for $^{104}$Ca+$\Ld$
systems vs. $f_s$. \label{fig2} }
\end{figure}

Fig.\ref{fig2} shows that the two-neutron separation energy of
$^{104}$Ca+$\Ld$'s systems increases from  the negative to the
positive with the rise of $n_\Ld$.   As the $n_\Ld$ increases, the
nucleon potential is enhanced by the attraction between hyperons
and nucleons, being able to fill in more neutrons under the
continuum.  The neutron drip line can thus be largely extended.
For instance, the drip line of Ca isotope can be extended from the
neutron number 50\cite{meng} to 82 as long as about 50 $\Ld$'s are
filled in. The dependence of the separation energy on the
isoscalar-isovector coupling $\Ld_{\rm v}$  reduces approximately
with the increase of $n_\Ld$. In Refs.\cite{schf1,schf2}, a large
class of MEMOs are constructed by adding multi-hyperons in stable
nuclei, while here we aim at the extension of the neutron drip
line by filling in $\Ld$ hyperons in unstable isotopes. As
multi-$\Lambda$ hypernuclei may exert their roles in extending the
neutron drip line and binding baryons more strongly than in normal
nuclei, they will potentially influence the astrophysical
nucleosynthesis and the abundances of elements. To search for
multi-$\Ld$ hypernuclei with the $n_\Ld$ up to 50 is possible but
certainly difficult\cite{jev}. Here, we just devote ourselves to
investigating the sensitivity of properties of such exotic systems
to the density dependence of the symmetry energy.

The sensitivity of the total binding energy to the $\Ld_{\rm v}$
is different for various $^{102}$Ca+$\Ld$'s systems. For
$^{102}$Ca+$\Ld$'s systems that are unstable against the neutron
emission at small $n_\Ld$ and unstable against the $\Ld$ hyperon
emission at very large $n_\Ld$, the total binding energies can be
modified by as much as tens of MeV  for  various $\Ld_{\rm v}$'s.
For stable $^{102}$Ca+$\Ld$'s systems whose binding energies are
situated on or very close to the minima, shown in Fig.\ref{fig1},
the modification to the total binding energies by various
$\Ld_{\rm v}$'s is quite small. For instance, the total binding
energy for $^{102}$Ca+50$\Ld$ in NL3 is just modified by up to
2MeV with various $\Ld_{\rm v}$ values.

A data to data correlation of  the neutron thicknesses between
$^{208}$Pb and stable multi-$\Ld$ nuclei can be established as in
Refs.\cite{todd,jiang1}. The neutron thickness of these stable
multi-$\Ld$ nuclei is very sensitive to the  $\Ld_{\rm v}$. For
instance, the uncertainty of the neutron thickness for
$^{102}$Ca+40$\Ld$ and $^{102}$Ca+50$\Ld$ is about 3 times that
for $^{208}$Pb.  The proton density distribution in these exotic
system is also sensitive to the $\Ld_{\rm v}$ and one third of the
uncertainty of the neutron thickness  comes from the protons. The
large uncertainty of the neutron thickness is attributed to the
addition of the $\Ld$ hyperons, which will be elucidated below.

\begin{figure}[thb]
\begin{center}
\vspace*{-25mm} \epsfig{file=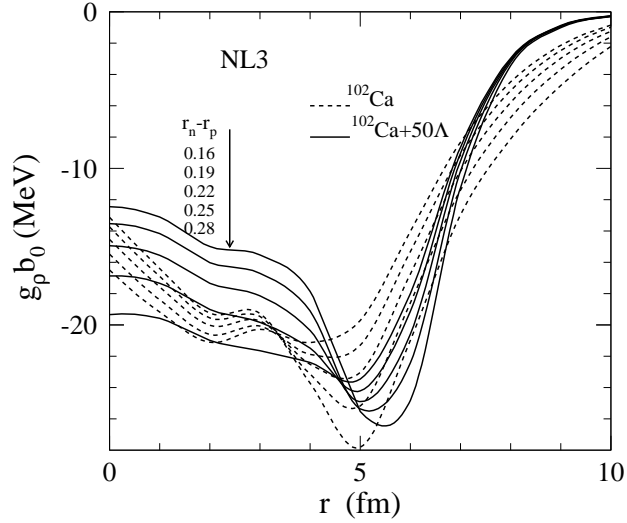,height=10.0cm,width=10.0cm}
\caption{ The isovector potential for $^{102}$Ca and
$^{102}$Ca+$50\Ld$ in NL3 vs. radius. The curves from above (near
the arrow) are obtained with various $\Ld_{\rm v}$'s  (equably
from 0.04 to 0.00) which give the neutron thicknesses  (in fm) in
$^{208}$Pb as denoted in ascending order. \label{fig4} }
\end{center}
\end{figure}

Usually, one modifies the isospin dependent potential considering
the shift of the isospin and the isovector field directly. Here,
we modify it by enhancing the isoscalar field ($\om$ field)
through filling in   $\Ld$'s without changing the isospin of the
system, which is fulfilled by considering the isoscalar-isovector
coupling. In Fig.\ref{fig4}, the isovector potentials are
displayed for $^{102}$Ca and $^{102}$Ca+$50\Ld$ in NL3. In S271,
similar results for these two systems can be obtained.  The
isovector potentials for various $\Ld_{\rm v}$'s in the exotic
isotope are largely shifted by filling in considerable $\Ld$
hyperons. The uncertainty of the isovector potential against the
$\Ld_{\rm v}$ in $^{102}$Ca is much larger in the larger radius
region, compatible with the understanding that $^{102}$Ca is
totally diffusive and unstable. Compared to results for
$^{102}$Ca, the sensitivity of the isovector potential to the
$\Ld_{\rm v}$ in $^{102}$Ca+$50\Ld$ is shifted to the central
region, and at the large radius the convergence occurs, which is
favored by the extension of the neutron drip line. The results
shown in Fig.\ref{fig4} for two exotic systems are related to
different density profiles. The central density in
$^{102}$Ca+$50\Ld$ is about one third higher than that in
$^{102}$Ca. The deviation of the symmetry energy for various
$\Ld_{\rm v}$'s enlarges  with the rise of the density beyond the
normal density or even much lower density\cite{brown,horo1,horo2}.
At the central region in $^{102}$Ca+$50\Ld$, the isovector
potential is more sensitive to the $\Ld_{\rm v}$  than that in
$^{102}$Ca. The isoscalar potential can also be shifted to a
certain degree by  the $\Ld_{\rm v}$, whereas for the normal
nuclei the shift is negligible. The sensitivity of the isovector
potential to the $\Ld_{\rm v}$ is responsible for the uncertainty
of the neutron thickness.

\begin{figure}[thb]
\begin{center}
\vspace*{-25mm} \epsfig{file=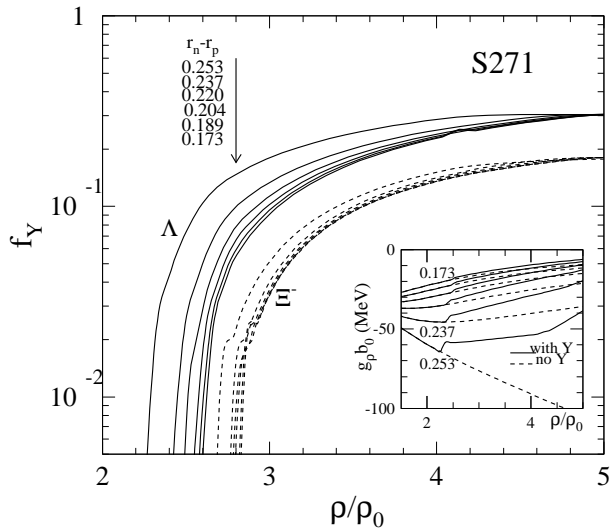,height=10.0cm,width=10.0cm}
\caption{ Hyperon fractions  in  S271 vs. baryon density in
hyperon-rich matter.  The  curves from above are for different
values of $\Ld_{\rm v}$ (equably from 0.00 to 0.05) that give the
neutron thicknesses in $^{208}$Pb as denoted in descending order.
The inset shows the isovector potential with hyperons and without.
\label{fig5} }
\end{center}
\end{figure}

The $\Ld$ hyperon constituent is important  in astrophysical bulk
matter to explore the effects of the density dependence of the
symmetry energy.  We plot the $\Ld$ and $\Xi^-$ fractions defined
by $f_Y=\r_Y/\r_B$ in Fig.\ref{fig5} with the S271 based on a
self-consistent calculation. The large sensitivity of the hyperon
fractions to the $\Ld_{\rm v}$ occurs at about $2\sim3\r_0$ where
the $\Ld$ hyperon starts to appear. Since the density as high as
$2\sim3\r_0$ is accessible through the heavy ion collisions with
the colliding energy about 1GeV per nucleon, the detection related
to the hyperons, especially the $\Ld$'s, in heavy ion collisions
is helpful to probe the density dependence of the nuclear symmetry
energy. The starting density for the $\Ld$ appearance is model
dependent. For instance, it is about 1.8$\r_0$ in NL3 and
2.2$\r_0$ in S271. As shown in the inset in Fig.\ref{fig5},  a
shift of the isovector potential starts to appear with the $\Ld$
appearance, and the isovector potential differs more from that
without hyperonization with the increase of the density. In
multi-$\Ld$ nuclei far from the normal drip line, the isovector
potential is shifted by filling in $\Ld$'s under the fixed
isospin, while here the shift comes from the change of the isospin
and  the isoscalar field together.  The convergent extent with
respect to the $\Ld_{\rm v}$ exists for the hyperon fraction in
the whole density domain. This indicates that it is favorable to
include the isoscalar-isovector coupling terms for the model
itself. At very high densities, the convergence of hyperon
fractions occurs. This is attributed to the fact that the isospin
is largely suppressed with the drop of the neutron fraction  at
high densities. The isovector potential is thus small compared to
the isoscalar potential, so that the isoscalar potential is little
changed through the isoscalar-isovector coupling. This implies
that the hyperstars\cite{schf3} whose core is solely composed of
hypermatter at  high densities will be insensitive to the
$\Ld_{\rm v}$.

\begin{figure}[thb]
\begin{center}
\vspace*{-25mm} \epsfig{file=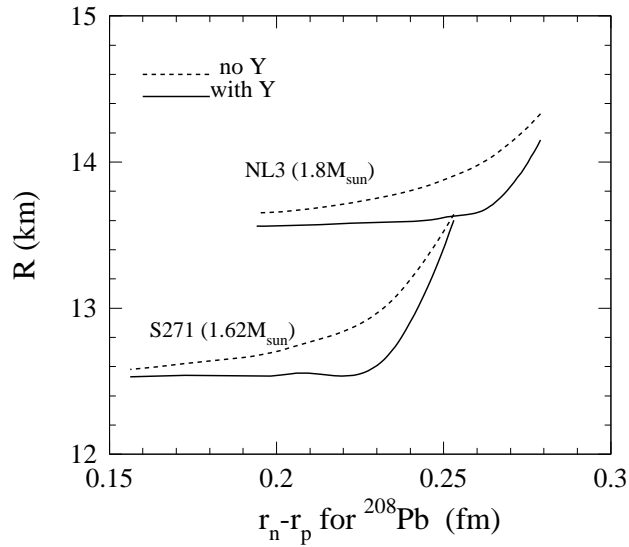,height=10.0cm,width=10.0cm}
\caption{ Radii of neutron stars as a function of the neutron
thickness in $^{208}$Pb.  \label{fig6} }
\end{center}
\end{figure}

The large reduction of the maximum mass  of neutron stars by
hyperonization, firstly pointed out by Glendenning (see
Ref.\cite{glend} and references therein), can be  reproduced here.
In NL3,  hyperonization  in the core reduces the maximum neutron
star mass by about 0.9$M_\odot$ from about 2.8$M_\odot$, and by
0.75$M_\odot$ from about 2.4$M_\odot$ in S271. Compared to models
in which hyperons are absent, hyperonization in the neutron star
whose mass is near the maximum mass increases the  central baryon
density by as much as 1$\r_0$ in the region $r<5$km. For a 1.4
solar-mass neutron star, the role of hyperonization is quite
limited because of relative low core densities. Using the
Oppenheimer-Volkoff equations for bulk matter in beta equilibrium,
we thus investigate neutron stars with relative large masses:
1.8$M_\odot$ with the NL3 model and 1.62$M_\odot$ with the S271
model.  In Fig.\ref{fig6}, the radii of neutron stars with
respective masses in two models are displayed as a function of the
neutron thickness for $^{208}$Pb. The radius R of  neutron star
without hyperonization increases with the neutron thickness in
$^{208}$Pb for a given parameter set, which is the same as in
Ref.\cite{horo2}. For neutron stars with hyperonization, the trend
is similar, whereas the correlation between the R and the neutron
thickness in $^{208}$Pb almost vanishes in a large region of the
neutron thickness in $^{208}$Pb. This property is consistent with
the convergent extent shown in Fig.\ref{fig5}. It indicates that
for a neutron star with hyperonization the radius R nearly just
depends on the on-off switch  rather than concrete values of the
isoscalar-isovector coupling in a given model. An accurate
measurement of the neutron radius for $^{208}$Pb at the Jefferson
Laboratory\cite{jeff} determines the density dependence of the
symmetry energy at low densities, as pointed out in
Ref.\cite{horo2}. The measurement of the radius of the neutron
star with hyperonization can discriminate two regions of  the
$\Ld_{\rm v}$'s to which the R is either very sensitive or
insensitive at high densities. Various information through
separate measurements may be combined together to  constrain the
isoscalar-isovector coupling that modifies the density dependence
of the symmetry energy in the whole density region.

In summary, we have studied the roles of the isoscalar $\Ld$
hyperons in exploring the density dependence of the symmetry
energy in  multi-$\Ld$ hypernuclei, hyperon-rich matter and
neutron stars with RMF models. Relationships between the
properties of three types of objects and the neutron thickness in
$^{208}$Pb are established with respect to the isoscalar-isovector
coupling that modifies the density dependence of the symmetry
energy. The exotic isotopes far from the neutron drip line can be
stabilized by filling in considerable $\Ld$ hyperons, and the
neutron drip line is substantially extended outwards. The
difference of the binding energy of multi-$\Ld$ hypernuclei from
different models is attributed to different symmetry energies. The
isovector potential together with the neutron thickness in
multi-$\Ld$ hypernuclei situated on the new drip line that is far
from the normal drip line is very sensitive to the
isoscalar-isovector coupling. In beta equilibrated hyperon-rich
matter, we have investigated the sensitivity of the hyperon
fractions to the isoscalar-isovector coupling based on the
self-consistent calculation for hyperons and nucleons. The large
sensitivity of the $\Ld$ hyperon fraction to the
isoscalar-isovector coupling occurs at about $2\sim3\r_0$. The
properties of neutron stars are investigated. An on-off effect
with respect to the isoscalar-isovector coupling exists for the
radius of neutron stars with hyperonization. Besides the detection
of $\Ld$ hyperons in the heavy ion collisions, the measurements of
the neutron star radius and neutron thickness in $^{208}$Pb could
help set up the constraints on the density dependence of the
symmetry energy.

\section*{Acknowledgement}
This work is partially supported by the National Natural Sciences
Foundation of China under grant No.10405031. The author thanks
Prof. M. Di Toro for helpful discussions.

\end{document}